# Why $T_c$ of MgB$_2$ is the highest in a number of diborides?

L.M. Volkova[1], S.A. Polyshchuk[1], F.E. Herbeck[2]

[1] Institute of Chemistry, Far Eastern Branch of Russian Academy of Science, 690022 Vladivostok, Russia

[2] Institute of Automation and Control Processes, Far Eastern Branch of Russian Academy of Science, 690041 Vladivostok, Russia.

**Abstract**
It is shown, that the problem rising $T_c$ diborides is a problem of nonstoichiometry in a plane of boron in diborides of heavy metals being compounds of variable composition.

**Comments:** The paragraph 2.4.2. from the chapter L.M. Volkova, S.A. Polyshchuk, F.E. Herbeck, "*Correlation of $T_c$ with Crystal Chemical Parameters in High-$T_c$ Cuprates, Diborides and Borocarbides: Concept of Arrangement and Function of Layered Superconductors*" in "FOCUS ON SUPERCONDUCTIVITY RESEARCH" Barry P. Martins., Ed., Nova Science Publisher, Inc. N.Y., 2003, (in print).

In [15] we examined a possibility of increasing $T_c$ in diborides on the ground of empirical correlation of $T_c$ with crystal chemical parameters in diborides ($T_c(J)$). Accoding to the calculation (Fig. 10):

- Among the diborides considered a superconductivity can be seen only in diborides W, Mo, Ru, Os and Ta, where $J<J_0$, and Mg, Cu(II), Ag(I) and Au(I), where $J>J_0$. The result obtained there suggest that the empirical absence or low-temperature superconductivity established in transition metal diborides with $J<J_0$ might be explained of presence B vacancies in B$_2$ plane. In the absence of vacancies in B$_2$ plane, $T_c$ of AB$_2$ diborides (A=W, Mo, Ru, Os) can be higher 77K, and in TaB$_2$ may treach 10K. The appearance of superconductivity and increasing $T_c$ values in NbB$_2$ and TaB$_2$ to need introduction of Nb(Ta)-vacancies or partial substitution of Nb$^{5+}$ (Ta$^{5+}$) on Al$^{3+}$, Ti$^{4+}$ or V$^{5+}$.
- Partial substitution of W, Mo, Ru and Os on more large cations with lower charge decreases $T_c$ with respect to one in initial diborides, but allows to conserve a stoichiometry in B$_2$ plane and to make a superconductors with high enough $T_c$.
- Partial substitution in MgB$_2$ of Mg on larger but with lower charge "cations" must heighten $T_c$.
- With pressure, $T_c$ must increase in superconductors with $J<J_0$ and decrease in ones with $J>J_0$.

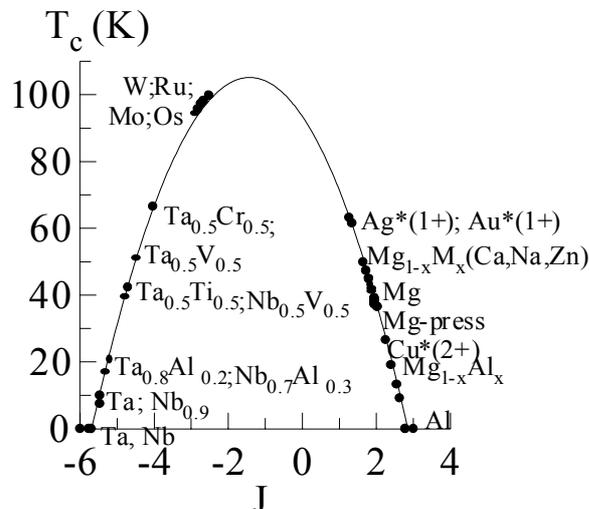

**Fig. 10.** $T_c$ as a function of $J$ in diborides AB$_2$: $T_c$ ($J$) of Egs. (7); $J$ calculated of Egs. (5) on the experimental (*theoretical) structural data for diborides.

Contrary to our forecasts about $T_c$ by stoichiometry in B plane and ideal structure it was not possible to highten $T_c$ in diborides. Only new superconductor - Nb-deficiency diboride with $T_c$ above 9K [127] confirms our outputs in [15]. According to calculations on Eqs. (7), $T_c$ of $Nb_{0.882}B_2$ ($a$=3.0987 A, $c$=3.3184 Å) is equal 9.1 K. It is necessary to understand, what is the reason of absence of a superconductivity or low temperatures of transition in W, Mo, Ru and Os diborides consists, which could be superconductors at $T_c$ higher 77K. Why the $T_c$ in $MgB_2$ is he highest among diborides? On the base of analysis of synthesis, character of change of structural parameters, and also the theoretical researches of a bond in diborides we came to certain conclusions, which we shall state below.

### 2.4.2. Problem of vacancies in a plane of boron

*Nature bond in diborides.* As a result of research of bonding nature in MgB2 is placed [154, 155], that within the boron planes the bond B-B are the distinct covalent, Mg atoms are strongly ionized and have no covalent bonding with atoms B, and between planes is delocalized a large amount of valence electrons, not participating in any covalent bonding. Thus, by [154, 155] in MgB2 there are two types of bonds: strongly covalent bonds within boron planes and delocalized, 'metallic-type' bonds between these planes. Many works research the bond ionicity in diborides. So, in [156] it is shown, that ionicity of A-B bonds decreases in the following order: Mg, Al, Mn, Y, Cr, Zr, Hf, Nb, Ta, V and Ti. The Mg-B bond has the largest value of ionicity of 96.8 % among these examined diborides.

*Crystal chemical dependences.* For plotting of crystal chemical dependences it was selected 52 of not substituted diborides $AB_2$ (A = Mg(II), Cu(II), Ag(III), Au(III), Al(III), Sc(III), Y(III), Lu(III), Cr(III), Mn(IV), Ti(IV), Zr(IV), Hf(IV), U(IV), Pu(IV), V(V), Nb(V), Ta(V), Mo(VI), W(VI), Os(VI) and Ru(VI)).

It was found, that in diborides there is no strict dependence of interatomic distances on a size of atom A. However, there is a tendency of increase of $d$(B-B) distances between boron atoms in a plane and $d(B_{pl}-A_{pl})$ distances between planes of boron atoms and atoms A with propagation of radius A (Fig 11 a, b) and, as a consequence, increase of $d(B_{pl}-A_{pl})$ with propagation $d$(B-B) distances (Fig. 11 c). In diborides examined the distances $d$(B-B) vary in limits 1.65 - 1.90 Å (Δ = 0.25 Å), $d(B_{pl}-A_{pl})$ in limits 1.43 - 2.00 Å (Δ = 0.75 Å). At approximately identical sizes of A atoms the distance $d$(B-B) and $d(B_{pl}-A_{pl})$ change, as a rule, in a random way. However, it is possible to select some features stipulated by A valence for example, in group where A = Mn(IV), Al(III), V(V), Ru(VI), Os(VI) and r = 0.670 - 0.685 Å, the distance between planes of metal and boron sharply enlarged (on 0.2 Å) in $AlB_2$ – the diboride with smaller A valence. But in two other groups A = Ti(IV), Cr(III) (r = 0.745 - 0.755 Å) and A = Hf(IV), Mg(II), Zr(IV) (r = 0.85 - 0.86 Å) this distance opposite diminished (on 0.1 Å in $CrB_2$ and on 0.06 Å in MgB2) for metals with less valence. It is impossible to explain these deviations by values of electronegativity of A atoms.

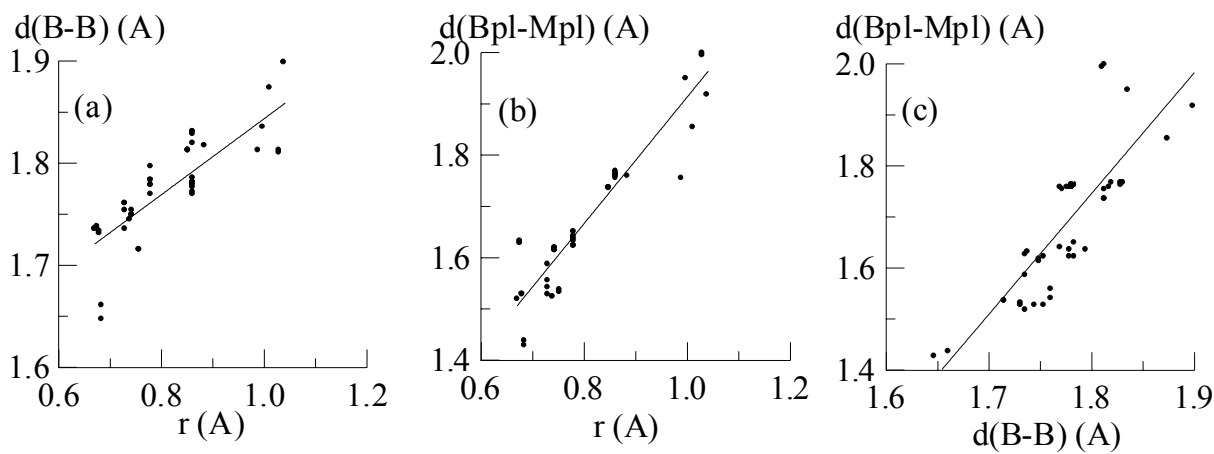

**Fig. 11.** $d$(B-B) as a function of r (a); $d(B_{pl}-A_{pl})$ as a function of r (b) and $d$(B-B) (c).

Such dependences are fulfilled best in limits of identical valence, if with increase of A radius its electronegativity decreases, for example, for A = V(V), Ta(V), Nb(V) or for A = Mn(IV), Ti(IV), Zr(IV), Pu(IV). Tervalent diborides the are divided into two subgroups with A = Al, Sc, Lu, Y and A = Cr, Ag, Au, having its own dependences. In group diborides of heavy metals Ru(VI), Os(VI), Mo(VI) and W (VI) there is a wide dispersion of distances.

*Variable composition.* The research of synthesis of diborides can shed light on this crystal chemical problem. In [153] it is proved, what even $MgB_2$ is not strictly stoichiometric compounds. There is in it an area of $Mg_{1-x}B_2$ homogeneity, and the parameters of a lattice vary in limits: $a$ = 3.065 - 3.091 Å and $c$ = 3.518 - 3.529 Å at change $T_c$ from 36.6 K [109] up to 39 K [79]. The maximal value of $T_c$ = 39 K is implemented in $Mg_{1-x}B_2$ with some deficit of Mg (x ~ 0.2). $T_c$ for stoichiometric $MgB_2$ is more than on 1.5 degrees [153] below.

The analysis of changes of structural parameters and $T_c$, obtained at substitutions of magnesium in $Mg_{1-x}B_2$ (A = Na, Ca, Zn, Cu, Ag) in [88, 123-125] for 0 < x≤0.2, shows, that they there are in the limits of parameter changes in not substituted $MgB_2$. Apparently, these changes grow out from nonstoichiometry of $MgB_2$, but not from effect of substitutions of magnesium. It is any doubt in results obtained at doping of $MgB_2$ by Al. It is explained in [123] by homogenous samples.

Striking example of nonstoichiometry of composition both planes of metal and boron are diborides of heavy metals. So, in [127] same $Nb_xB_2$ (x = 1 - 0.52) is synthesized and it is shown, that with increase of Nb contents the parameter of lattice $a$ is incremented, and parameter $c$ varies with a random way. If to select from that number the samples with improved composition (x = 0.68, 0.76 and 1.08), it is possible to see, that with increase of Nb contents the parameter $a$ is incremented (on 0.01 Å), and the parameter $c$ is reduced (on 0.05 Å). The molybdenum diborides are notable for the deficit of atoms of boron, which can reach 40% [137]. Parameters of a lattice in $MoB_{2-x}$ [137] vary with an opposite way by change of parameters in a number borides ($Nb_xB_2$) with metal deficit: increase of boron contents of from 60 % up to 69.5 at. % the parameter $a$ decreases from 3.041 Å up to 3.005 Å, and the parameter $c$, opposite is incremented from 3.060 Å up to 3.173 Å.

These two processes in diborides of heavy metals, apparently, are interdependent: the increase of concentration of metal concerning equilibrium one result in inevitable for the occurrence of vacancies in a plane of boron, and on the contrary.

Addition of 4% Zr in molybdenum diborid of [126] allows to shift this equilibrium and to increase a boron contents in the system $(Mo_{0.96}Zr_{0.04})B_y$. It confirms by change of properties of compounds in this system. For $(Mo_{0.96}Zr_{0.04})_xB_2$ for 1.0≥ x≥ 0.85 $T_c$ increases from 5.9 to 8.2 K with the introduction of metal vacancies.

In our opinion, in [126] there are no weighty proofs of complete absence of vacancies in a plane of boron, as occupancy of the B site were not improved by neutron diffraction. The increase of the lattice parameter $a$ together with cutting of the parameter $c$ by increase of metal contents in the system $(Mo_{0.96}Zr_{0.04})_xB_y$ can indicate both on increase of metal content, and on increase of vacancies in boron plane. Moreover, the parameter $c$ (3.128 Å) in stoichiometric $(Mo_{0.96}Zr_{0.04})B_2$ [126] is less on 0.01 Å, than in samples MoB2-x [137] with boron deficit of 30.5%. And as is shown above, the parameter c in $MoB_{2-x}$ is incremented with increase of boron contents. Practically it is possible only to reduce the number of vacancies in boron plane, as for example, it is made in [126]. Theoretically it is possible to propose and other method of obtaining of superconducting borides. It is necessary to remove half of atoms of boron by saving diboride structure and ordering of vacancies in boron plane. For formation of necessary carriers concentration it is possible to enter the vacancies into a metal plane or to substitute a part of metal atoms on atoms with less valency.

Thus, we have shown, that the deviations from stoichiometry are characteristic for diborides, therefore does not exist well-defined dependence between of crystal chemical parameters (Fig. 11). The existence of delocalization of valence electrons between layers [154] and various types of bonds is, apparently, a reason of variably composition of these compounds. The inappreciable deviations from stoichiometry are observed only in MgB2 because of the greater ionicity degree of Mg-B bond among diborides. Despite of indirectness of an ionic bond, the first coordination sphere of each Mg ion should be saturated practically with identical number of opposite ions for saving an electroneutrality of a crystal. For this reason the area of $MgB_2$ homogeneity is not great, in a plane of boron there are no vacancies as in diborides of heavy metals and $T_c$ is the highest.

Most variable of composition appears by decrease of a ionicity degree of bond in heavy metals diborides. It was shown on an example of Nb and Mo diborides. It is known, that the nature of deviations from stoichiometric composition in binary of variable composition is, that in structure of a crystal there are defects of various types. The characteristic elementary defect is the occurrence of vacancies in a sublattice of one of component ($A_{1-x}B_2$ or $AB_{2-x}$) or simultaneously in sublattices of two components ($A_{1-x}B_{2-y}$). In the latter case, the combination of nonstoichiometry in two planes of metal and boron can give the mutual compensation of their influence, therefore the violations of stoichiometry can theoretically not be.

The researches of high-$T_c$ cuprates showed, that the defects in $CuO_2$ plane result in suppression of a superconductivity. Therefore, the problem of occurrence of a superconductivity and rise $T_c$ in diborides of heavy metals, such as W, Mo, Ru and Os is a problem of obtaining of diborides without vacancies in a boron plane. As the presence of defects is incorporated in a nature of these compounds of variable composition to reach the highes possible $T_c$ it will be not possible.